\documentclass[aps,prd,twocolumn,superscriptaddress]{revtex4}
\pdfoutput=1

\usepackage{graphicx} 
\usepackage{amssymb}
\usepackage{amsmath}
\usepackage{amsthm}
\usepackage{slashed}
\usepackage[colorlinks]{hyperref}
\usepackage{color,rotating}
\usepackage{bbm}
\usepackage{array}
\usepackage{pstricks}
\usepackage{pstricks-add}
\usepackage{colonequals}
\usepackage[utf8]{inputenc}

\DeclareMathAlphabet{\EuRoman}{U}{eur}{m}{n}
\SetMathAlphabet{\EuRoman}{bold}{U}{eur}{b}{n}

 \graphicspath{{./figures/}}

\def\di{\displaystyle}

\def\bg{\begin{eqnarray}\begin{array}{rcl}\displaystyle}
\def\eg{\end{array} &\di    &\di   \end{eqnarray}}
\def\bm#1{\begin{eqnarray}\begin{array}{#1}\di}
\def\bmo#1{\begin{eqnarray*}\begin{array}{#1}\di}
\def\bml#1#2{\begin{eqnarray}\begin{array}{#1}\label{#2}\di}
\def\bgo{\begin{eqnarray*}\begin{array}{rcl}\displaystyle}
\def\ego{\end{array} &\di    &\di \nonumber  \end{eqnarray*}}

\def\btensor#1#2{\renew\left#1\begin{array}{#2}\di}
\def\brtensor#1#2#3{\ren#3\left#1\begin{array}{#2}}
\def\botensor#1#2{\renew\left#1\begin{array}{#2}}
\def\etensor#1{\end{array}\right#1}

\def\s0#1#2{\mbox{\small{$ \frac{#1}{#2} $}}}
\def\0#1#2{\frac{#1}{#2}}

\def\s{\sigma}

\def\ren#1{\renewcommand{\arraystretch}{#1}}

\def\renew{\renewcommand{\arraystretch}{1}}

\newcommand{\UV}{{\small UV}}
\newcommand{\IR}{{\small IR}}
\newcommand{\FRG}{{\small FRG}}

\newcommand{\QCD}{{\small QCD}}

\newcommand{\LPA}{{\small LPA}}
\newcommand{\PDE}{{\small PDE}}
\newcommand{\WKB}{{\small WKB}}

\allowdisplaybreaks

\begin{document}

\title{Solving functional flow equations with pseudo-spectral methods}

\author{J. Borchardt}
\email[]{julia.borchardt@uni-jena.de}
\affiliation{Theoretisch-Physikalisches Institut, Universit\"at Jena,
Max-Wien-Platz 1, 07743 Jena, Germany}
\author{B. Knorr}
\email[]{benjamin.knorr@uni-jena.de}
\affiliation{Theoretisch-Physikalisches Institut, Universit\"at Jena,
Max-Wien-Platz 1, 07743 Jena, Germany}


\begin{abstract}
We apply pseudo-spectral methods to integrate functional flow equations with high accuracy, extending
earlier work on functional fixed point equations \cite{Borchardt:2015rxa}. The advantages of our
method are illustrated with the help of two classes of models: first, to make contact with literature, we investigate flows of the
O$(N)$-model in 3 dimensions, for $N=1, 4$ and in the large $N$ limit. For the case of a fractal dimension, $d=2.4$, and $N=1$, we follow the flow
along a separatrix from a multicritical fixed point to the Wilson-Fisher fixed point over almost 13 orders of magnitude.
As a second example, we consider flows
of bounded quantum-mechanical potentials, which can be considered as a toy model for Higgs inflation. Such flows
pose substantial numerical difficulties, and represent a perfect test bed to exemplify the power of pseudo-spectral
methods.
\end{abstract}

\maketitle

\section{Introduction}

A lot of fundamental problems in particle physics, many-body systems, or even the quantisation of gravity, arise in situations where coupling constants can grow large.
Standard methods as perturbation theory fail to describe these cases, as they are based on an
expansion in powers of a small quantity. As not every theory can be simulated straightforwardly by means of discretisation, we shall follow another path here, namely the functional 
renormalisation group (\FRG{}), which is a continuum method based on the Wilsonian idea of integrating out
momentum modes successively. In particular, this work makes use of the formulation of the exact
renormalisation group by Wetterich \cite{Wetterich:1992yh}, which has been applied to a wide range of systems, {\it{e.g.}}
scalar field theories \cite{Tetradis:1993ts,Berges:2000ew,Delamotte:2007pf,Morris:1997xj,Bervillier:2007rc,Litim:2010tt,Benitez:2011xx,
Codello:2012ec,Percacci:2014tfa,Codello:2014yfa},
fermionic systems \cite{Metzner:2011cw,Scherer:2010sv,Gies:2010st,Braun:2010tt,Braun:2011pp,Gies:2014xha,Boettcher:2013kia},
critical phenomena \cite{Kopietz:2010zz,Janssen:2012pq,Gies:2009da,Scherer:2013pda,Litim:2002cf,Benitez:2009xg,Jakubczyk:2014isa},
gauge theories \cite{Reuter:1993kw,Gies:2006wv,Pawlowski:2005xe,Litim:1998qi,Braun:2011iz,Tripolt:2013zfa,Braun:2014ata,Mitter:2014wpa}
and quantum gravity \cite{Reuter:1996cp,Niedermaier:2006wt,Percacci:2007sz,Manrique:2009uh,Benedetti:2010nr,Benedetti:2012dx,Eichhorn:2015bna,Demmel:2014hla,
Christiansen:2012rx,Falls:2013bv,Christiansen:2014raa,Folkerts:2011jz,Eichhorn:2011pc,Harst:2011zx,Dona:2013qba,Percacci:2015wwa,Gies:2015tca,Gies:2016con}.

Technically, when applying the \FRG{} to a given theory, one has to solve a coupled system of
non-linear (in general integro-)differential equations. Only a few cases are known where one can
find analytic solutions. In all other cases, the system is considered in a subspace of the space spanned
by all operators allowed by symmetry, and the resulting equations are then solved numerically.

Plenty of information can already be retrieved from the fixed point structure of the theory. As an example,
the fixed point structure in a condensed matter system may characterise the phase diagram, and the eigenvalues
of the perturbations around the fixed points give the critical exponents, controlling {\it{e.g.}} the scaling behaviour
near phase transitions. We recently put forward a method to numerically tackle functional fixed point equations
globally and with very high precision \cite{Borchardt:2015rxa}.

Not all questions can be answered by studying fixed points alone. The full functional flows need to
be solved, {\it{e.g.}}, in regions of physical interest when all couplings run fast, or for the analysis of
first order phase transitions. 
In this work, we extend the ideas of \cite{Borchardt:2015rxa} to solve
flow equations with the help of pseudo-spectral methods.
In order to benchmark our method, we investigate models which are well understood or widely studied
in the \FRG{} context, or can also be controlled by other techniques.
First, we study flows of the O$(N)$-model in various dimensions and near as well as away from criticality.
Then we investigate quantum mechanical examples with bounded or non-analytic potentials, as exact results
from directly solving the Schr\"odinger equation can be calculated and compared to.
Further applications of the method can be found in 
\cite{Heilmann:2014iga,Borchardt:2016xju}.

We emphasise that the methods presented here are heavily used in other contexts \cite{Boyd:ChebyFourier,Robson:1993}, as e.g. finding solutions to
Einstein’s equation \cite{Ansorg:2003br,Macedo:2014bfa}.
First applications to \FRG{} problems have been given in \cite{Litim:2003kf,Fischer:2004uk,Gneiting:2005,Zappala:2012wh}.
Additionally, let us point out that full functional flows were already solved in the past employing finite element or finite difference methods \cite{Berges:1996ja,Berges:2000ew,Adams:1995cv,
Papp:1999he,Bohr:2000gp,Schaefer:2004en,Bonanno:2004pq,Caillol:2012zz,Roscher:2015xha,Boettcher:2014tfa,Fischbacher:2012ib,Synatschke:2010jn,Mitter:2013fxa,Strodthoff:2011tz,Herbst:2013ufa,Pelaez:2015nsa}.

This paper is organised as follows: 
In \autoref{sec:PSM}, we review some aspects of pseudo-spectral methods and
apply them to study concrete functional flow equations. 
Then, \autoref{sec:FRG} gives a short overview
of the \FRG{}, including the truncation and the resulting flow equations that we solve. 
Consequently, \autoref{sec:ONmodel} deals with interesting flows of the O$(N)$ model,
especially for large $N$, $N=1$ and $N=4$ in $d=3$ or $d=2.4$ dimensions.
Afterwards, \autoref{sec:QM}
discusses both analytical and numerical results on three quantum mechanical potentials which are bounded both from
below and above. Finally, \autoref{sec:sum} contains a short summary.

All numerical results presented here were obtained with C++ code, using the libraries {\small{BOOST}} \cite{boost},
Eigen \cite{eigen} and Blitz \cite{BlitzLib}, and the 80-bit data type \textit{long double}.

\section{Pseudo-spectral methods and their implementation}\label{sec:PSM}

This section elucidates the advantages of pseudo-spectral methods, and how we apply them
to solve flow equations. Pseudo-spectral methods in general are based on orthogonal polynomials. We will
focus on Chebyshev polynomials of the first kind here, which are defined by
\begin{equation}
 T_n(\cos(x)) = \cos(n x) \, ,
\end{equation}
for all natural numbers $n$. A given well-behaved function $f$, defined on the interval $[a,b]$, can thus be expanded as
\begin{equation}
 f(x) = \sum\limits_{i=0}^\infty a_i T_i\left(2\frac{x-a}{b-a}-1\right) \, .
\end{equation}
The usefulness of Chebyshev polynomials in general is (at least) threefold:
\begin{itemize}
 \item for analytic functions, an expansion in Chebyshev polynomials converges exponentially fast,
 \item evaluations of the interpolant at any point are easily accessible (by the Clenshaw algorithm) and
 derivatives of the function are also computable with a minimum amount of effort (similar to the Clenshaw algorithm),
 \item the highest retained coefficient in an approximation provides a good error estimate for the accuracy of such an expansion.
\end{itemize}
More details on all of these points are collected in \cite{Borchardt:2015rxa}, a general overview of
Chebyshev polynomials can be found in {\it{e.g.}} \cite{Boyd:ChebyFourier}.

For the remainder of this section, let us consider a \PDE{} of one function $f$ in 2 variables,
\begin{equation}
 \mathcal L[f(x,t)] = 0 \, ,
\end{equation}
where $x\in I_x$ and $t \in I_t$ with 
$I_x,I_t \subseteq \mathbb{R}$, and $\mathcal L$ denotes an in general non-linear (pseudo-)differential operator.
In the case of flows of the O$(N)$-model, we specialise $I_x=[0,x_\text{max}]$ because no boundary effects occur.
By contrast, if the potential is bounded as in \autoref{sec:QM}, boundary effects emerge 
which can only be avoided by taking the function $f$ on the whole positive axis $I_x=[0,\infty)$ into consideration.
In particular, we will compactify in $x$ direction \footnote{In \cite{Borchardt:2015rxa} we used rational
Chebyshev functions to resolve the asymptotic part. From the point of view of an implementation in an
actual code, it is more convenient to work with a compactification as presented here. This does not have
any impact on convergence, so any choice made is purely conventional.}.
The specific form of the compactification that we employ is
\begin{equation}\label{eq:compactification}
 \bar x = \frac{x}{1+x} \, .
\end{equation}
Although not used here, it is worth mentioning that
for unbounded functions which grow like $x^n$ for $x\to\infty$,
it is useful to compactify as
\begin{equation}
 \bar f = \frac{f}{(1+x)^n} \, .
\end{equation}
These transformations in combination have the special virtue that they map
polynomials to polynomials: for example, if
\begin{equation}
 f(x,t) = \sum\limits_{n=0}^{N_x} a_n(t) x^n \, ,
\end{equation}
after transformation to the new coordinates, we have
\begin{align}
 \bar f(\bar x,t) &= \frac{1}{\left(1+\frac{\bar x}{1-\bar x}\right)^{N_x}} \sum\limits_{n=0}^{N_x} a_n(t) \left(\frac{\bar x}{1-\bar x}\right)^n \notag \\
 &= \sum\limits_{n=0}^{N_x} a_n(t) \bar x^n (1-\bar x)^{{N_x}-n} \, ,
\end{align}
which is polynomial in $\bar x$. For the sake of readability, from now on
we will drop any bar on transformed quantities when there is no danger of confusion.

\begin{figure}
 \includegraphics[width=0.8\columnwidth]{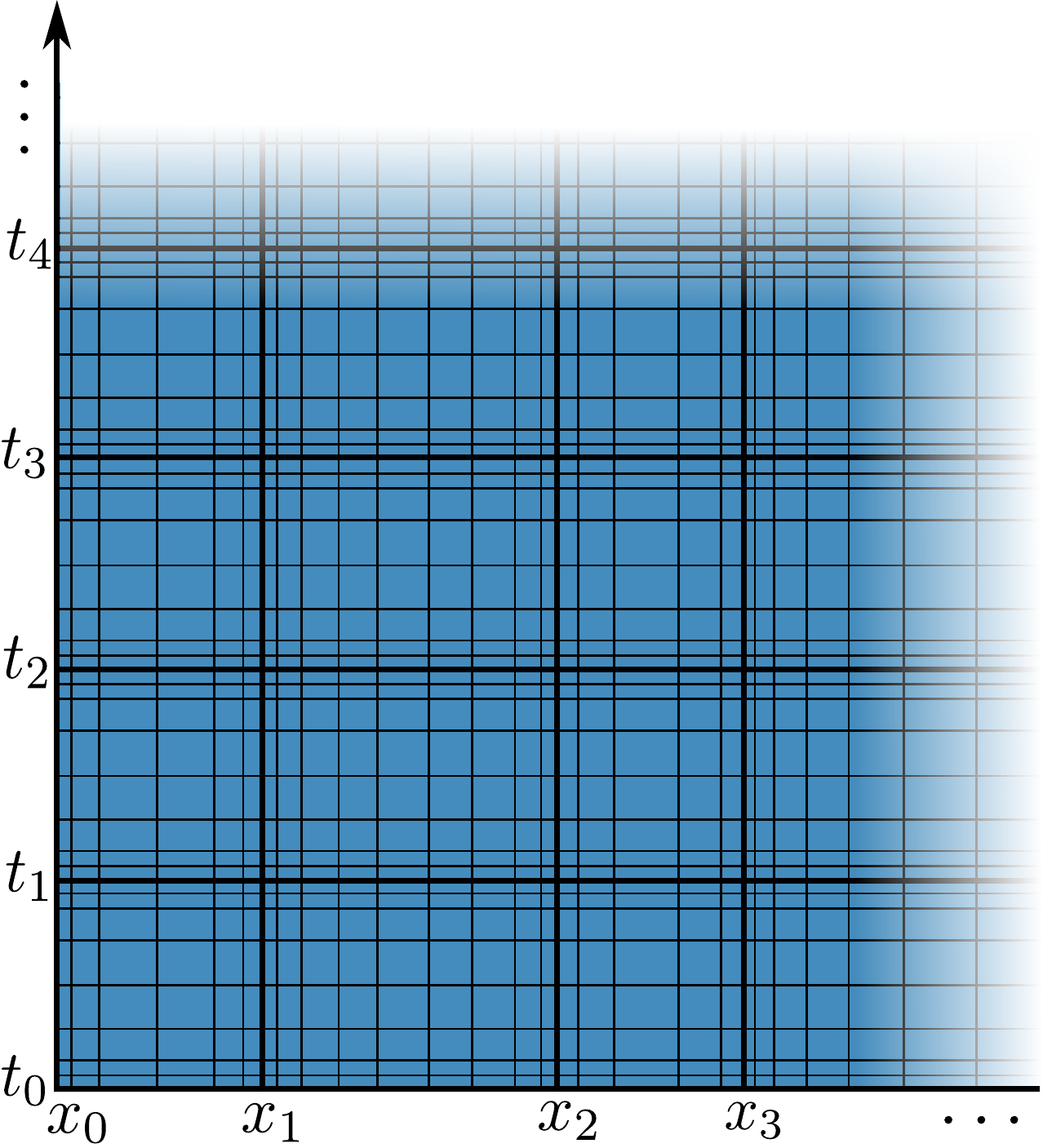}
 \caption{Sketch on the decomposition of the domain of definition into subdomains.
 The $t_i$ and $x_i$ denote the boundaries of the subdomains in $t$ and $x$ direction, respectively.
 For each subdomain an own expansion of $f$ is used.
 The thin grid lines depict the Chebyshev grid for each subdomain.}
 \label{fig:multipledomain}
\end{figure}

In any case, to obtain high efficiency and better convergence, it is useful to decompose the $x$ and $t$ domain of definition into subdomains.
This is illustrated in \autoref{fig:multipledomain}.
The \PDE{} is solved simultaneously on all $x$ subdomains of a specific $t$ slice $t \in [t_i,t_{i+1}]$.
Imposing the initial condition on the first slice and continuity on the following slices
leads to an initial value problem on every time patch.
For that purpose, we expand the (potentially compactified) function $f$ on every $t$ slice (mapped to $[-1,1]$),
in every (compactified) $x$ subdomain (also mapped to $[-1,1]$) as a tensor product,
\begin{equation}
 f(x,t) = \sum\limits_{i=0}^{N_x} \sum\limits_{j=0}^{N_t} a_{ij} T_i(x)T_j(t) \, .
\end{equation}
This ansatz can be inserted into the \PDE{}. To solve for the coefficients $a_{ij}$, we employ
the collocation method, where the \PDE{} is evaluated at a set of collocation points and the subsequent algebraic,
non-linear system of equations is solved by a stabilised Newton-Raphson iteration.
As collocation points, we specify a one-sided Radau grid in $t$ direction, including the endpoint of the patch,
and a Gauss grid on all $x$ subdomains. 
As already mentioned, one has to match the initial condition in $t$ direction
as well as the function value and a certain number of derivatives at the boundaries of the subdomains in $x$ direction. The 
exact number of conditions is dictated by the order of the differential equation: if the order is $p$, then
$p-1$ derivatives have to be matched.

Lastly, let us remark that a generalisation to multiple functions is straight-forward by introducing
a tensor product expansion for any given function. In the context of specific truncations,
one may additionally deal with functions that do not depend on $x$, such as single running couplings. These are naturally incorporated
by taking $N_x=0$.

\section{Functional Renormalisation Group}\label{sec:FRG}

The functional renormalisation group is a nonperturbative tool
for successively integrating out quantum fluctuations in a controlled way.
For this, an effective average action $\Gamma_k$ is introduced which smoothly connects
the microscopic action $\Gamma_\Lambda = S_\text{cl}$ at the ultraviolet (\UV{}) cutoff $\Lambda$
and the macroscopic action $\Gamma_0 = \Gamma$.
This flow is described by the Wetterich equation \cite{Wetterich:1992yh},
\begin{equation} \label{eq:Wetterich}
 \partial_t \Gamma_k = \frac{1}{2} \mathrm{STr}\left[ \left(\Gamma^{(2)}_k + R_k \right)^{-1} (\partial_t R_k) \right], \quad t=\log\bigg(\frac{k}{\Lambda}\bigg),
\end{equation}
that is an exact functional, integro-differential equation.
Here, $\Gamma^{(2)}_k$ denotes the second functional derivative of $\Gamma_k$ with respect to the fields
and the super-trace $\mathrm{STr}$ stands for summation (integration) over discrete (continuous) indices and
provides a minus sign for Grassmann-valued fields, \textit{i.e.} fermions.
The regulator function $R_k$ prevents the flow from divergencies both in the \UV{} and infrared (\IR{}).
For more information on the \FRG{}, we refer the reader to, \textit{e.g.}
\cite{Berges:2000ew,Pawlowski:2005xe,Gies:2006wv,Kopietz:2010zz,Delamotte:2007pf}.

The common case is that equation \eqref{eq:Wetterich} can only be solved within a certain truncation of the effective average action.
A systematic expansion is provided by the derivative expansion.
The systems we consider in what follows go all back to the same ansatz for the effective average action,
\begin{equation} \label{eq:effaction}
 \Gamma_k[\sigma] = \int \mathrm{d}^d x \left\{ \frac{1}{2} Z\, (\partial_\mu \sigma^a)(\partial^\mu \sigma^a) + U(\sigma^a \sigma^a/2) \right\}.
\end{equation}
Within the local potential approximation (\LPA{} or \LPA{}'), the full field and scale dependence of the effective potential $U$
is retained.
The wave function renormalisation $Z$ is field independent, and is therefore only a function of the scale $k$ (\LPA{}')
or constant during the flow (\LPA{}).
In \autoref{sec:ONmodel} the index of the bosonic scalar field $\sigma$ counts the $N$ different components.
When we consider quantum mechanical systems in \autoref{sec:QM}, $\sigma$
stands for the position $x$, and the index counts the space dimensions.
Also, the integration and differentiation in the action are w.r.t. the Euclidean time coordinate, $\mathrm{d}x \equiv \mathrm{d}\tau$.
In both cases the invariant $\rho$ is given by $\rho=\sigma^a \sigma^a/2$.

\section{Flows of the O\texorpdfstring{$(N)$}{(N)}-model}\label{sec:ONmodel}

The O$(N)$-symmetric model is a relevant model for many aspects of particle physics, condensed matter systems and \QCD{}.
It consists of bosonic fields only, exhibits a global symmetry, there is an analytical solution for the large $N$ limit 
(see \cite{Tetradis:1993ts,Tetradis:1995br} for approaches using the Wetterich equation), and 
the $N=1$ case is even exactly solvable in $d=1$ \cite{Isingref1} and $d=2$ \cite{PhysRev.65.117}.
Additionally, it provides physically interesting features for $d<4$, such as a rich fixed point structure.
Therefore, it is a good testing ground for demonstrating properties of pseudo-spectral methods.

The flow of the first derivative of the potential with respect to the field is numerically more stable than the flow of the potential itself \cite{Hasenfratz:1985dm}.
Therefore, we employ the flow equation for the dimensionless quantities
$u'(\tilde\rho)=\partial_{\tilde\rho}u(\tilde\rho)$ \cite{Tetradis:1993ts} where $\tilde\rho = Z k^{2-d}\rho$ and $u = U/k^d$,
\begin{align}\label{eq:flowON}
 \partial_t u'(\tilde\rho) = &(-2+\eta)u'(\tilde\rho)+(d-2+\eta)\tilde\rho u''(\tilde\rho) \notag \\ 
 &- \frac{4 v_d}{d}\left( 1-\frac{\eta}{d+2} \right)\times \\
 &\times\bigg( \frac{3u''(\tilde\rho)+2\tilde\rho u'''(\tilde\rho)}{(1+u'(\tilde\rho)+2\tilde\rho u''(\tilde\rho))^2}
 +\frac{(N-1)u''(\tilde\rho)}{(1+u'(\tilde\rho))^2}\bigg) \notag\,.
\end{align}
Here, $v_d^{-1} = 2^{d+1}\pi^{d/2}\Gamma(d/2)$, and the anomalous dimension is defined as $\eta=\partial_t \ln Z$,
and given by \footnote{Here, we use the Goldstone anomalous dimension, even for $N=1$, as this
leads to superior results.}
\begin{equation}\label{eq:flowZ}
 \eta = \frac{16 v_d}{d}\frac{\tilde\rho_0 u''(\tilde\rho_0)^2}{(1+2\tilde\rho_0 u''(\tilde\rho_0))^2} \, ,
\end{equation}
evaluated at the vacuum expectation value (vev) $\tilde\rho_0$.
For the regularisation the linear optimised regulator is employed \cite{Litim:2001up}, $R_k(p^2) = Z (k^2 - p^2)\theta(k^2-p^2)$.
For aspects of optimisation, see also \cite{Litim:2000ci}.
In the first part of this section we set $\eta\equiv0$ ($Z\equiv1$) which becomes exact for large $N$.
In the last part, we take the scale dependence of the wave function renormalisation into account.

\subsection{Flows for \texorpdfstring{$d=3$}{d=3} and at large \texorpdfstring{$N$}{N}: A comparison}

In order to demonstrate the power of pseudo-spectral methods on a specific example, 
we compare the analytical flow for large $N$ with the numerically computed one.
For that purpose, we choose trajectories in the symmetry broken phase close to criticality
to show stability of the numerical method for $6$ orders of magnitude ($t\in [0,-12.4]$).
We use \eqref{eq:flowON} in the limit $N\rightarrow \infty$ \cite{Tetradis:1995br} (where one only retains the scaling part
and the fluctuation part proportional to $N$) and 
switch to dimensional quantities as soon as the vev starts scaling exponentially in $t$.
We expand the first derivative of the potential on $[0,0.2]$ for the dimensionless and on $[0,0.2k_\text{S}]$
for the dimensional flow, where $k_\text{S}$ is the scale of switching between both regimes.

The initial condition reads
\begin{equation}
U'_{\Lambda}(\rho) = -0.008443603515625+0.5\rho 
\label{eq:initialcond}
\end{equation}
at $t=0$ or $k=\Lambda$, where $\Lambda$ is the \UV{}-cutoff. 
All dimensional quantities are to be understood in units of $\Lambda$, which we set to $1$.
For switching to the dimensional version of \eqref{eq:flowON}, we choose $t_\text{s}=\ln (k_\text{S}/\Lambda)=-10.1$.
Furthermore, the temporal subdomains and $N_t$ are taken to achieve exponential convergence
down to machine precision in this direction.
In order to compare the analytical potential \cite{Tetradis:1995br} with the numerically computed one, 
we employ the maximum norm of their difference as error criterion.

\begin{figure}
 \includegraphics[width=0.98\columnwidth]{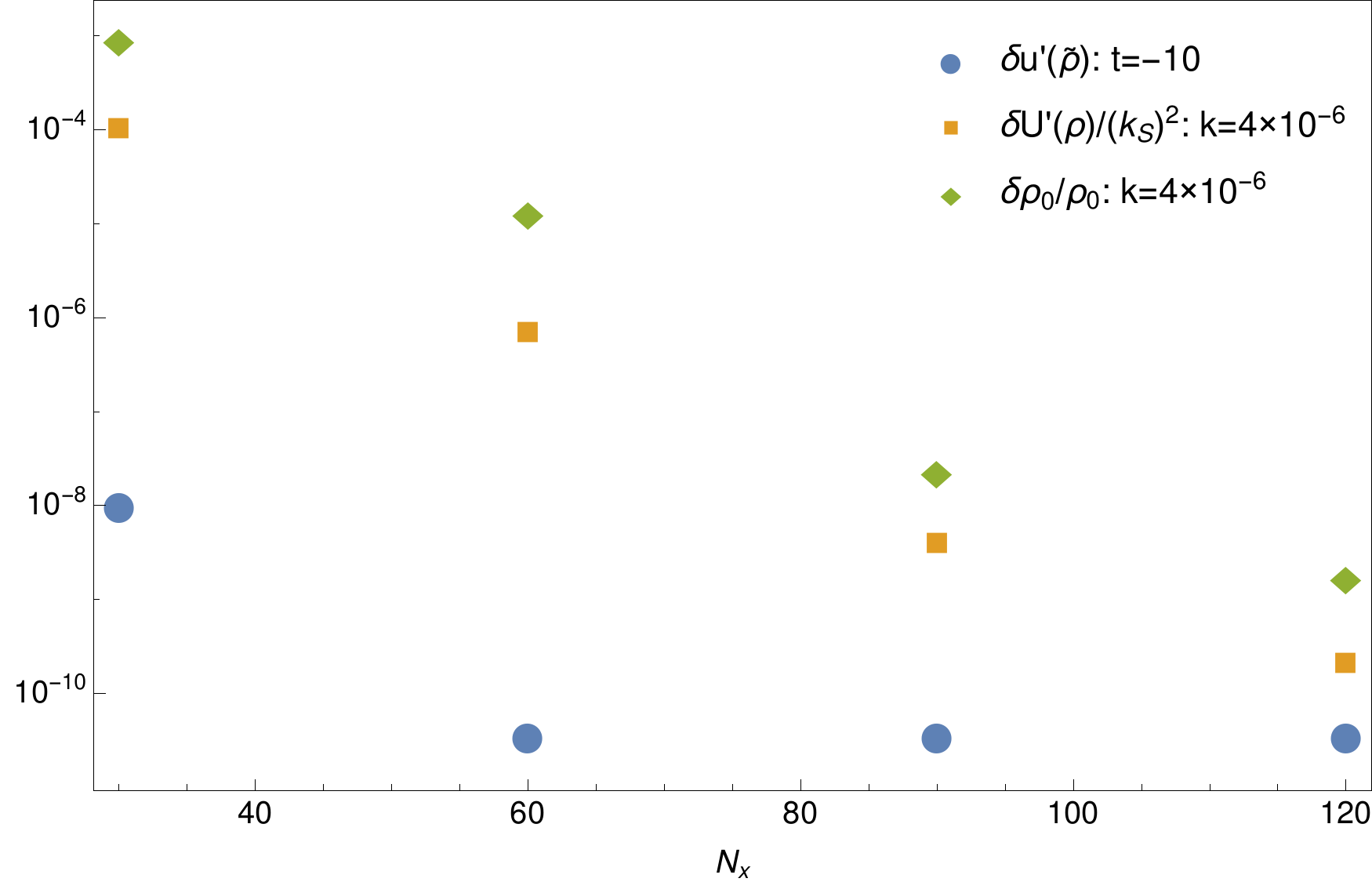}
\caption{Absolute and relative error ($\delta u'(\tilde \rho)$,$\delta U'(\rho)$ and $\delta\rho_0/\rho_0$) of the first derivative of the potential and the vev, respectively,
as a function of the number of coefficients $N_x$ in field direction.
The errors $\delta U'(\rho)$ and $\delta\rho_0/\rho_0$ decreases exponentially.
For the error of $u'(\tilde \rho)$ at $t=-10$, one can see a plateau 
which is due to the condition of the differential equation. 
This indicates that the solution is accurate to almost machine precision.
}
\label{fig:largeNerror}
\end{figure}

In \autoref{fig:largeNerror} the absolute deviation of the numerical flow from the analytical one 
in dependence of the number of the coefficients $N_x$ in field direction can be seen.
The flow was compared at two scales: $t=-10$ ($k=4.5 \cdot 10^{-5}$), before switching to dimensional quantities, and $k=4\cdot10^{-6}$  ($t=-12.4$), after switching to dimensional quantities,
where we have stopped the integration.
We also depict the relative error of the vev at this scale.
The more coefficients are taken into account, the higher the accuracy, which can be seen by
the exponential convergence of $\delta U'(\rho)$ and $\delta\rho_0/\rho_0$ in particular.
For the error $\delta u'(\tilde \rho)$ at $t=-10$ we see a plateau for $N_x\gtrsim 60$.
This can be explained by the condition of the differential equation.
To illustrate this, we compare two analytically computed solutions,
one with the initial condition \eqref{eq:initialcond}, and the other with a small deviation from it.
To obtain an error of about $\sim 10^{-11}$ at $t=-10$,
one can allow for a deviation of $10^{-18}$ for the constant term,
and $10^{-16}$ for the linear term,
which is about the order of magnitude that we can resolve with \textit{long double}.
This example indicates how carefully time integration has to be done for staying close to the original trajectory.
On the other hand, it shows that we have integrated out the flow close to machine precision over many orders of magnitude for $N_x\gtrsim 60$.
This fact is supported by the exponential convergence till $\sim10^{-18}$ of the coefficients.

For the \IR{} flow, the decrease of the error is slower, but still tends to the lower bound $\sim 10^{-11}$ for a large number of coefficients.
The error is now dominated by the truncation error of the expansion of the potential in field direction since convexity starts to set in.
From the asymptotic decrease of the last coefficients for $N_x\gtrsim 60$, we obtain a measure for the truncation error
which agrees very well with the errors depicted in \autoref{fig:largeNerror}.
It is based on an estimate for the sum over the neglected coefficients.
In order to achieve machine precision, more coefficients are needed.

We conclude that in a large part of theory space, the pseudo-spectral flow is highly efficient, and we generically observe
exponential convergence for an increasing number of Chebyshev coefficients. Therefore, we concentrate in the following on the most
challenging part of theory space involving the built-up of non-analyticities, the first adumbration of which we just started to discuss.

\subsection{Flows for \texorpdfstring{$d=3$}{d=3} and \texorpdfstring{$N=1,\,4$}{N=1,4}}

\begin{figure*}
 \includegraphics[width=\columnwidth]{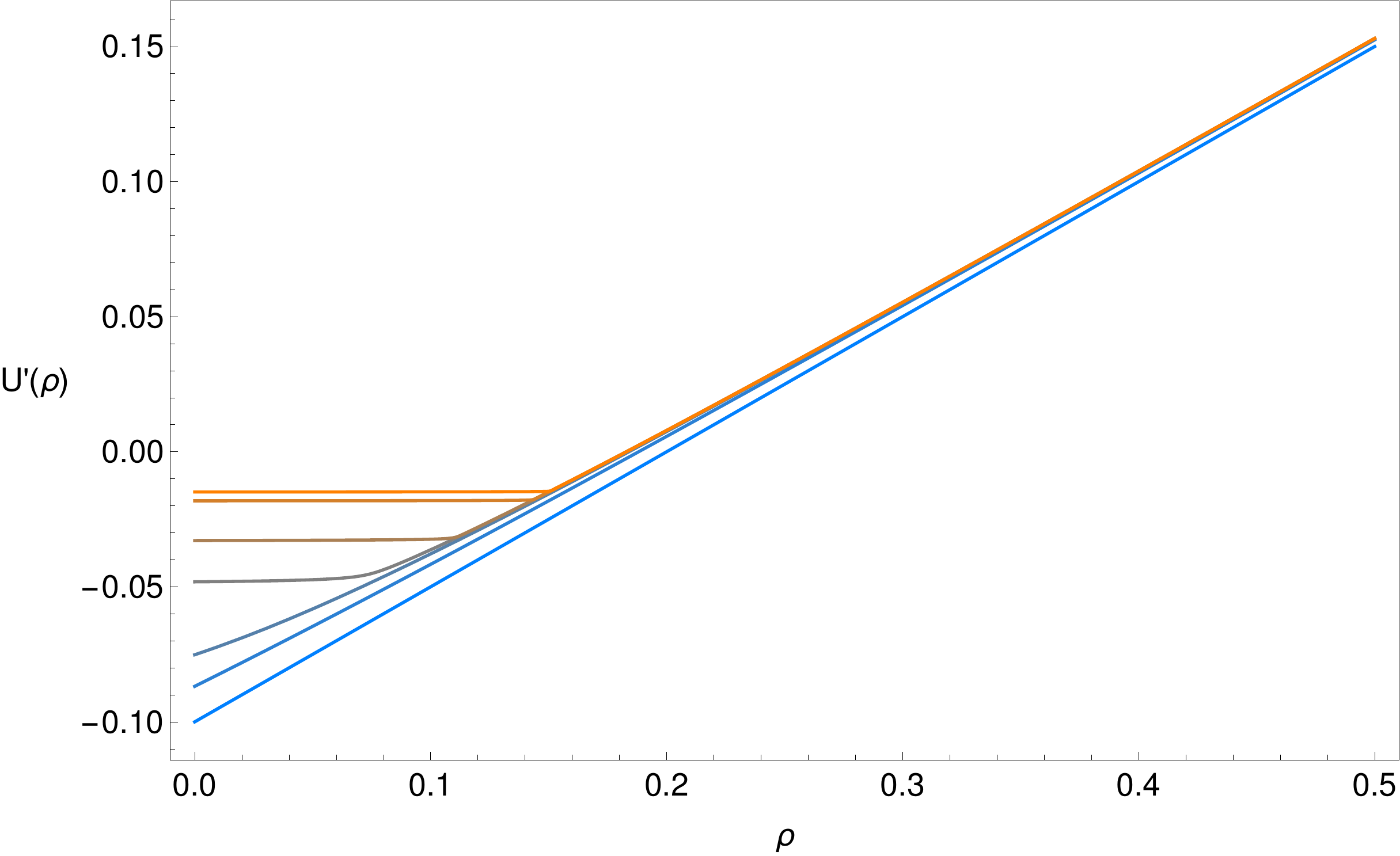}
 \includegraphics[width=1.01\columnwidth]{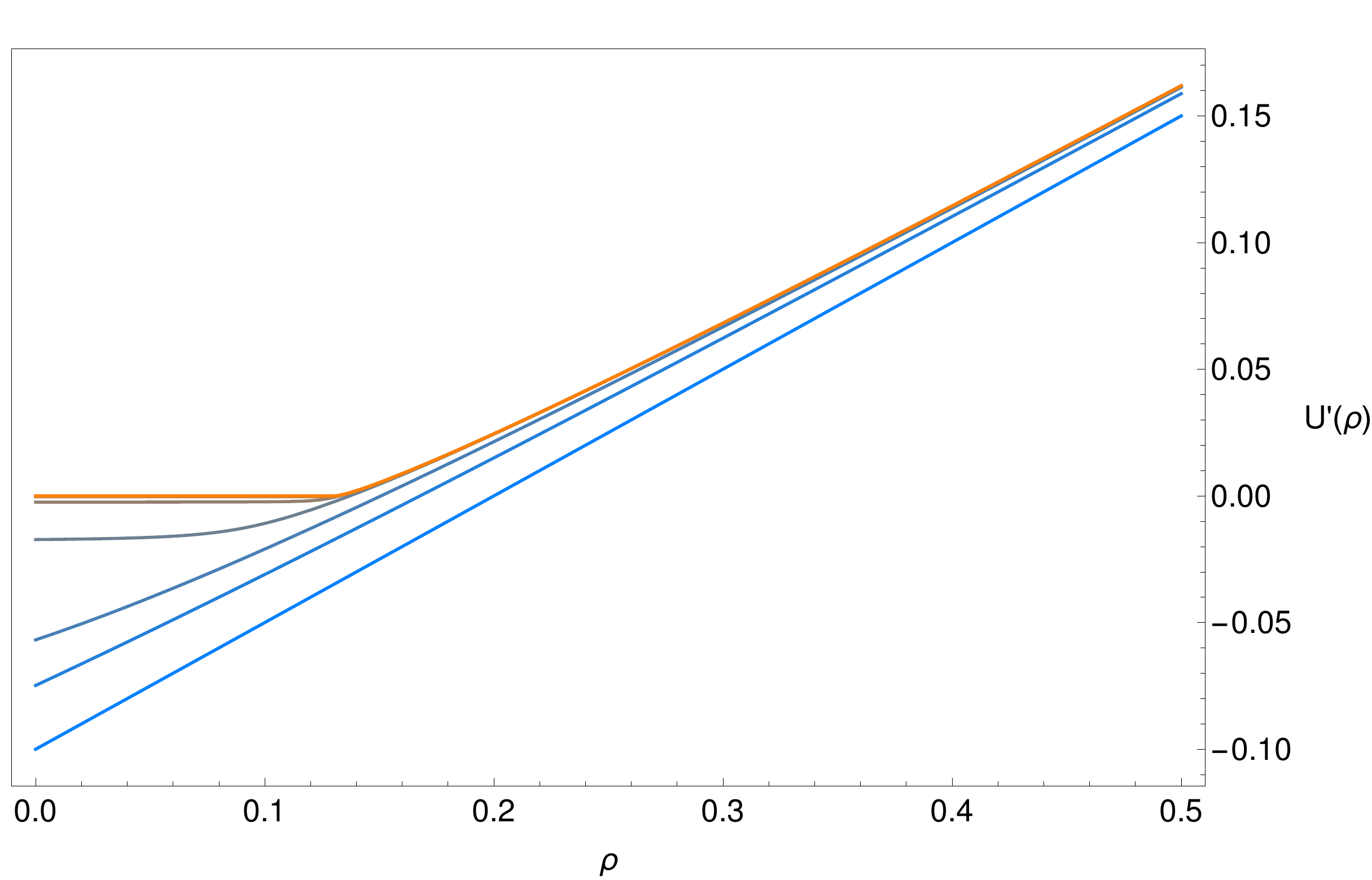}
 \caption{Evolution of $U'(\rho)$ from blue (bottom) to orange (top) for $N=1$ (left panel; $t=0,-0.5,-1,-1.5,-1.7,-2,-2.1$)
 and $N=4$ (right panel; $t=0,-0.5,-1,-2,-3,-4,-5,-13$). Convexity is seen in the flattening of $U'(\rho)$ for small fields $\rho<\rho_0$.
 Whereas $U''(\rho)$ is still continuous for $N=4$, in the single scalar case a jump occurs.}
  \label{fig:EvolutionN=1,4}
\end{figure*}

\begin{figure}
 \includegraphics[width=\columnwidth]{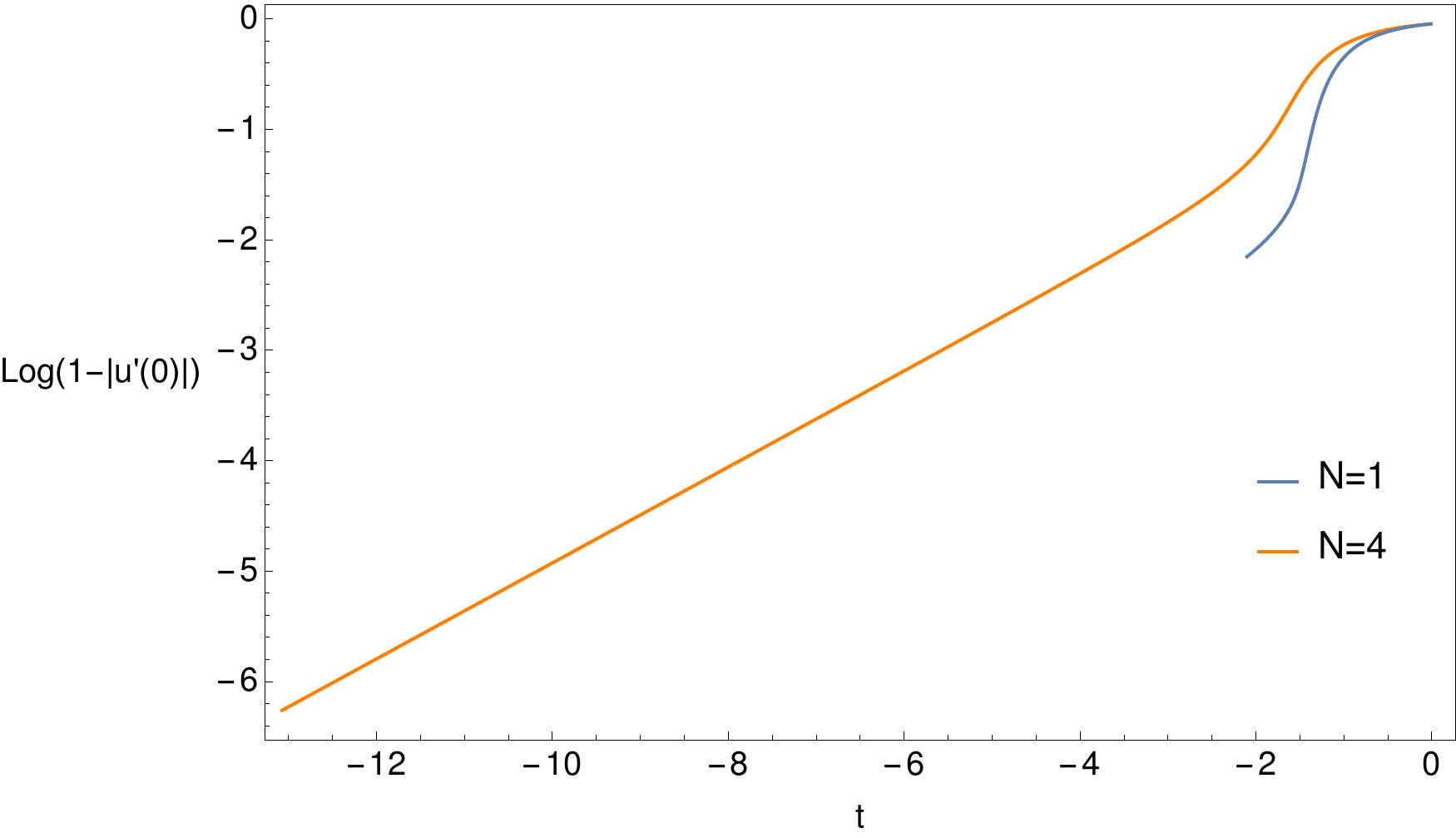}\\
 \caption{$u'(0)$ approaches the singularity $-1$ for $t\rightarrow -\infty$.
 Due to the stronger non-analyticity in the single scalar case, the numerically computed 
 flow ceases to exist earlier.}
 \label{fig:Singularity}
\end{figure}

In the spontaneously symmetry broken phase, the effective potential is nonconvex for all intermediate scales $k>0$.
On the other hand, it is known that the effective potential has to be convex at $k=0$ even in \LPA{} \cite{Berges:2000ew,Litim:2006nn}. 
While the outer region already is convex, the inner region becomes flat during the \IR{} flow.
Since the radial mass does not vanish for $N=1$, the curvature jumps at the vev at $k=0$.
By contrast for $N>1$, the influence of Goldstone bosons partly suppresses this non-analyticity.
The propagators $\propto (1+u'(\tilde\rho))^{-1}$ and $\propto (1+u'(\tilde\rho)+2\tilde\rho u''(\tilde\rho))^{-1}$ 
flow towards the singularity for small $\tilde\rho$,
pushing the convexity mechanism forward.

We picked out two particular values for $N$, namely $N=1$ and $N=4$.
The following calculations are done with the dimensional version of \eqref{eq:flowON} 
since we choose the initial condition to be far from criticality, $U'_{\Lambda}(\rho) = -0.1+0.5\rho$, at $k=\Lambda$.
It is convenient to use the logarithmic time scale $t$ instead of $k$.
After a few orders of magnitude dimensional scaling can be observed.

\autoref{fig:EvolutionN=1,4} depicts the evolution of $U'(\rho)$ for $N=1$ and $N=4$,
from large to small scales.
The approach to convexity is clearly visible. The built-up of the corresponding non-analyticity can be monitored over a range of scales,
especially for $N=4$.
As $U'(\rho)$ for $N=1$ has an edge at $\rho_0$ at $k=0$ where $U''(\rho_0)$ jumps, the flow is numerically much harder to track
and finally breaks down earlier.
The reason is as follows:
Exponential convergence of the coefficients is only guaranteed if the function is analytical.
For $k=\Lambda$, the convergence of the coefficients in field direction is very fast. 
Plateaus that build up for higher order coefficients are on the level of the machine precision.
However, for low scales $k$, the requirement for exponential convergence is not fulfilled anymore.
Thus, we observe a slower convergence of the coefficients till it breaks down.
Although this problem cannot be avoided completely, there are two possibilities for improvement:
On the one hand, one can simply take more coefficients.
This will not cure the problem completely since the convergence becomes too slow and finally, an unacceptably large number of coefficients is needed.
On the other hand, one can choose the domains in such a way that the non-analyticity lies close to the boundary of two neighbouring domains.
For that reason, we have used $24$ or $16$ domains for $N=1$ or $N=4$, respectively.
The high accuracy of pseudo-spectral methods prevents the flow to jump over the singularity of the propagator for a long time.
\autoref{fig:Singularity} shows how the flow approaches the singular point.
Due to the reasons given above, for $N=4$ we get closer to $u'(0)=-1$ in comparison to $N=1$.

We have shown that pseudo-spectral methods can also be applied to numerically challenging problems, such as convexity.
Let us emphasise that the convergence of the expansion coefficients is strongly connected to the properties of the solution.
Therefore, it is not surprising that the numerical effort increases the closer the singularity is approached.
In contrast to other approaches adjusted to tackle convexity issues \cite{Pelaez:2015nsa,Bonanno:2004pq},
we again point out that pseudo-spectral methods have a striking advantage:
The error is controllable by the convergence pattern of the expansion coefficients, which was especially demonstrated in the previous section.
Furthermore, if only \IR{} quantities are of interest, \textit{e.g.} the vev, 
they can be inferred from the flow before convexity becomes challenging.
We obtain $\rho_0 = 0.183$ for $N=1$ and $\rho_0 = 0.130$ for $N=4$ and the radial mass $m^2_\text{R} = 2\rho_0 U''(\rho_0) = 0.168$ for $N=1$.
It is worth mentioning that the vev for $N=4$ deviates by $2\%$ from the vev derived from the analytical large $N$ solution.
That indicates that the large $N$ limit already is a proper approximation for the $N=4$ case.

Finally, note that pseudo-spectral methods are easily extendable to higher truncations, 
\textit{e.g.} taking a field-dependent wave function renormalisation or $p^4$ operators into account \cite{Heilmann:2014iga}.

\subsection{Flow between two criticalities for \texorpdfstring{$N=1$}{N=1}}

\begin{figure*}
 \includegraphics[width=\columnwidth]{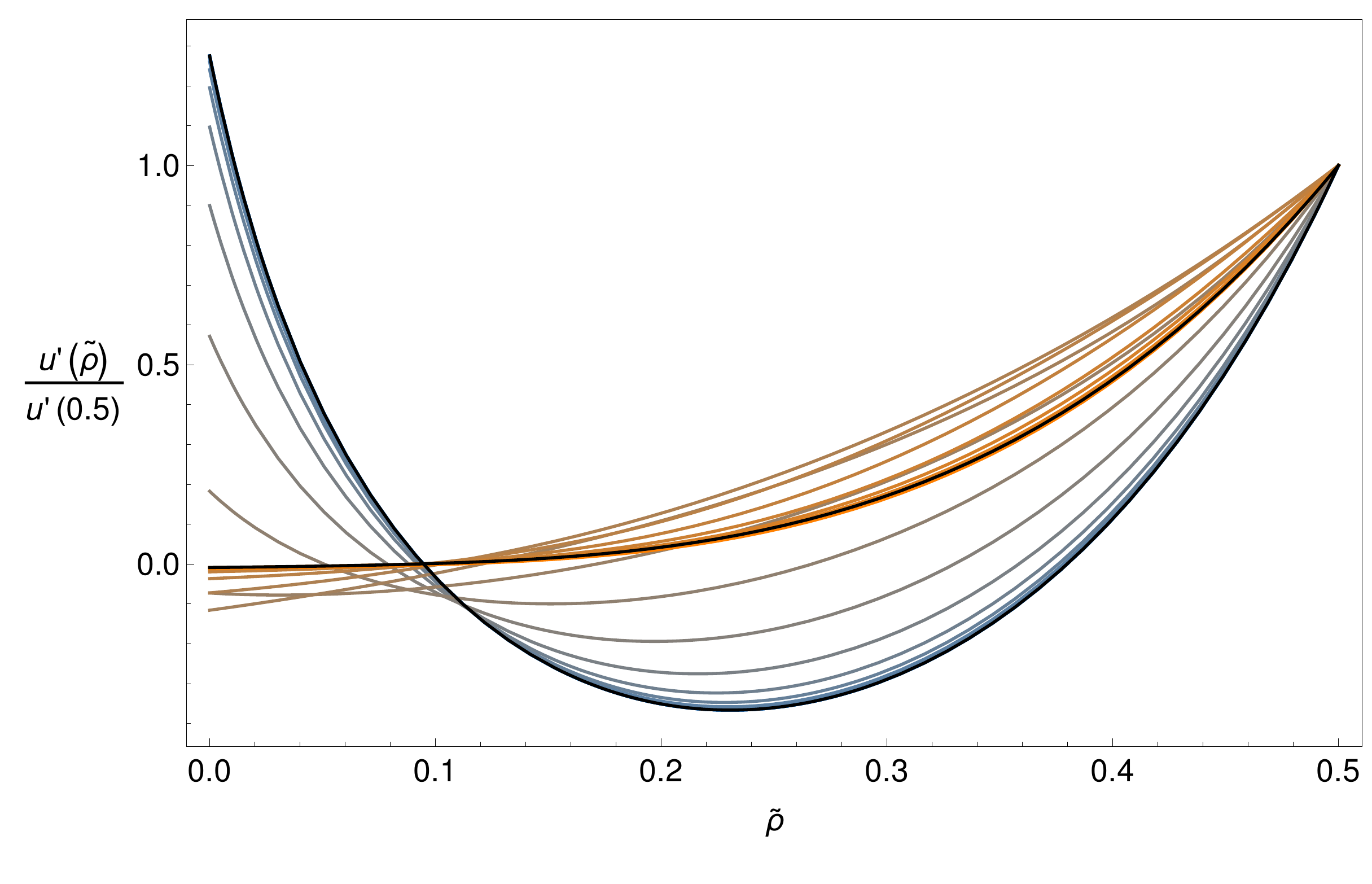}
 \includegraphics[width=\columnwidth]{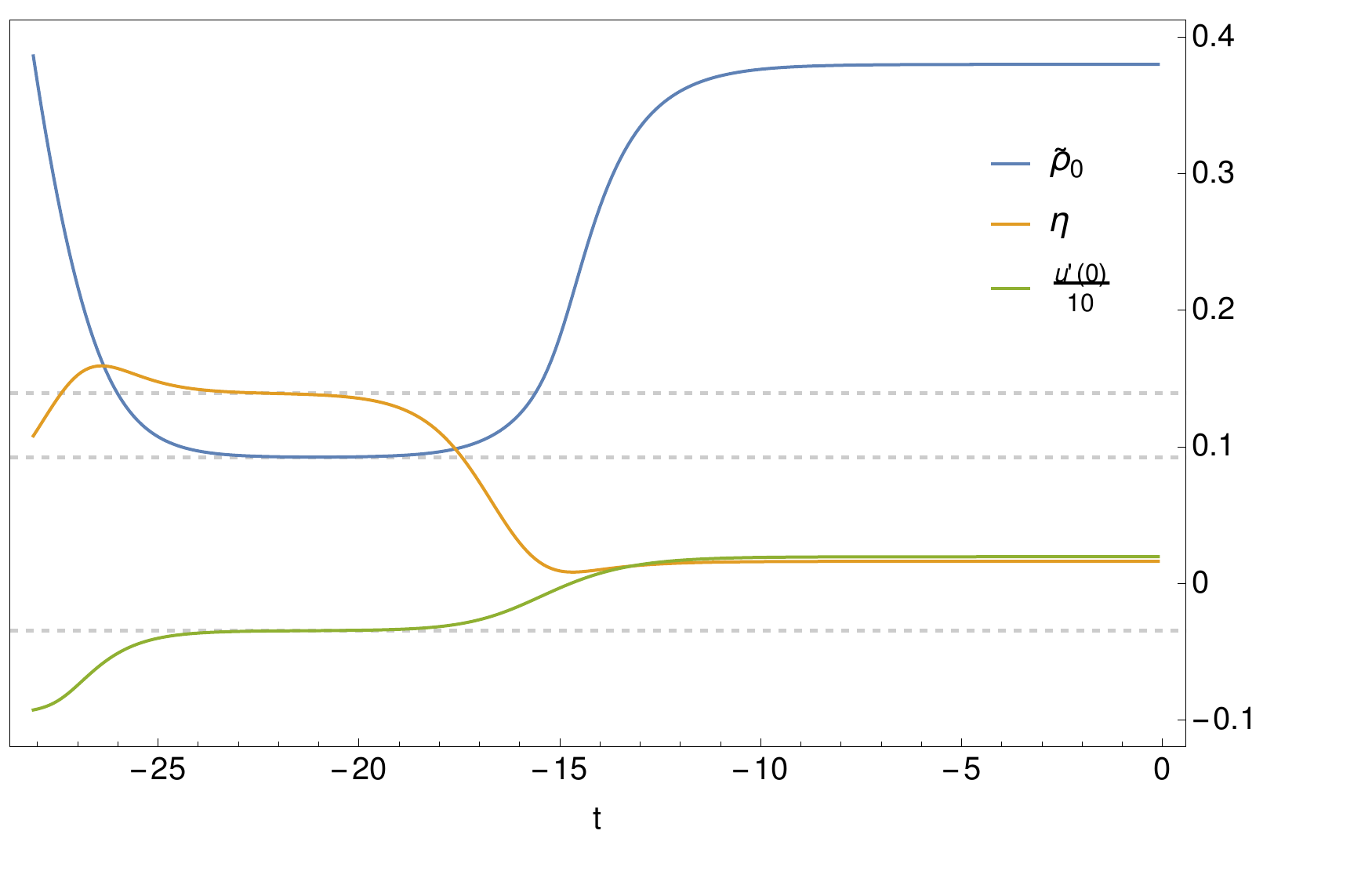}
\caption{Flow between two criticalities. Left panel: Flow from the tricritical fixed point potential (blue) to the Wilson-Fisher potential (orange), $t\in [0,-25]$.
 The fixed point potential computed from the fixed point equation are depicted as well (black). Right panel: Flow of the anomalous dimension $\eta$, the vev and $u'(0)$.
 The grey dashed lines denote the values of the Wilson-Fisher fixed point solution obtained from solving the fixed point equation.}
\label{fig:Flowcrit}
\end{figure*}

In the previous section, we have investigated flows far from criticality.
However, for $d<4$ nontrivial fixed points occur.
The first one is the well known Wilson-Fisher fixed point.
Lowering the dimension further, multicritical fixed points emerge at certain critical dimensions $d_{c,i} = 2i/(i-1)$ for $i\geq3$.
This is discussed in \cite{Codello:2012ec,Codello:2012sc,Codello:2014yfa} in detail.
In \cite{Borchardt:2015rxa} global solutions of the first four fixed point potentials for $d=2.4$ are given.
Now, we take a closer look to the first two fixed points, the Wilson-Fisher fixed point among them, in $d=2.4$.
We are interested in a trajectory connecting both (separatrix).
Therefore, we start at the tricritical fixed point with a small deviation constructed from a linear combination of its relevant eigenperturbations.
For our calculations we employ \eqref{eq:flowON} and \eqref{eq:flowZ} with the wave function renormalisation being scale dependent.
As initial conditions we use the results of \cite{Borchardt:2015rxa}.

For approaching the Wilson-Fisher fixed point during the flow, we have to fine-tune the linear combination of both relevant directions of the tricritical fixed point.
The perturbation is mainly along the second relevant (subleading) direction.
The flow strongly depends on the numerical parameters.
This is not surprising since small perturbations in the relevant direction may lead to large deviations during the flow as already seen for the large $N$ case.
\autoref{fig:Flowcrit} shows the deformation of the potential $u'(\tilde\rho)$ from the tricritical fixed point to the Wilson-Fisher fixed point during the flow.
The inner minimum of the tricritical fixed point potential disappears.
In the right panel the anomalous dimension, the vev and $u'(0)$ are plotted over the logarithmic scale.
Whereas all quantities and the potential itself stay at the tricritical fixed point for many orders of magnitude, they finally approach the Wilson-Fisher fixed point.
This can be seen from the plateaus at $-17\gtrsim t\gtrsim-25$.
The relevant direction becomes irrelevant at the Wilson-Fisher fixed point.
Finally, the flow carries the critical behaviour of the Wilson-Fisher fixed point although we have started at the tricritical fixed point.
We emphasise that for such flows a very stable numerical method is indispensable for which pseudo-spectral methods are well suited.

\section{Quantum Mechanics with a bounded potential}\label{sec:QM}

In this section we present results on the energies of the ground and first excited states
of a selection of three quantum-mechanical potentials
obtained by solving the flow equation for the derivative of the effective potential.
This is specifically suited to test our methods, as a direct comparison with
other methods and the exact answer is possible, and in the \FRG{} framework, an extension
to quantum field theory is straightforward.

In particular, we will focus on potentials that are bounded both from below and above.
Physically, such potentials are interesting, {\it{e.g.}}, in the 
context of Higgs inflation \cite{Saltas:2015vsc}. Technically, the flows of such potentials necessitate a global resolution
 - if the flow of only a finite region in $x$ is considered, one encounters boundary
effects that destabilise the flow.
To put the results in perspective, we will compare them
with the (numerically) exact values, as well as values obtained from various analytic approximations.

\begin{figure*}
\includegraphics[width=\textwidth]{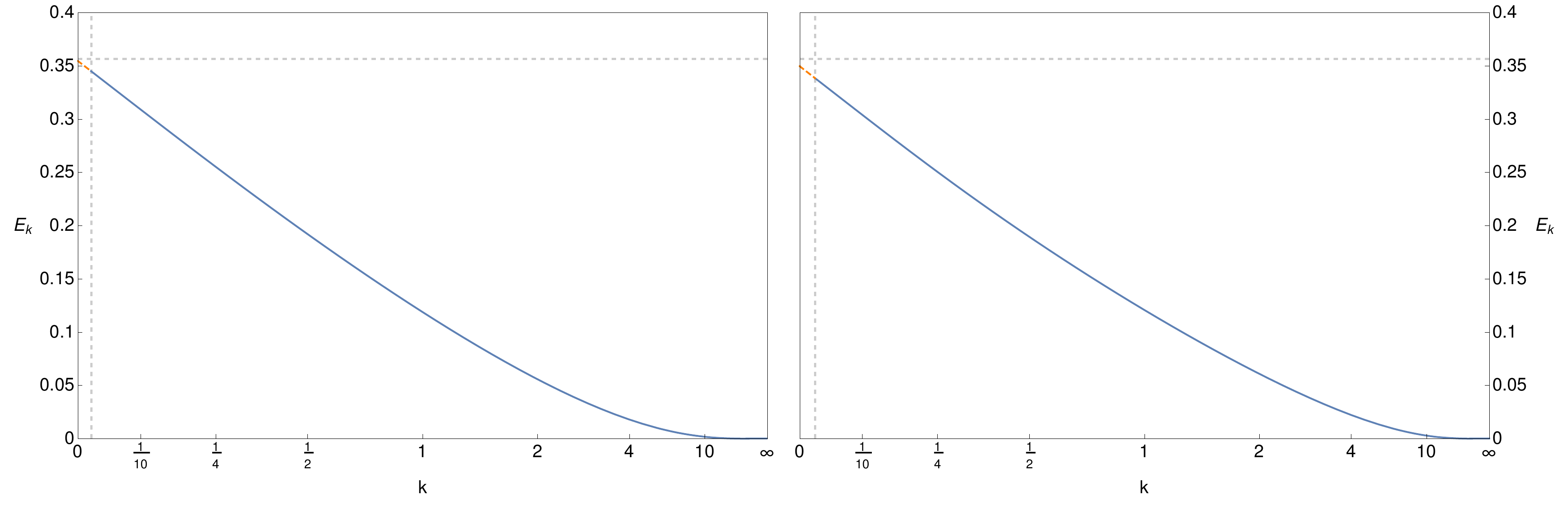}
\caption{Flow of the effective potential at vanishing position, which gives the effective ground state energy at
scale $k$, $E_k$, for both Callan-Symanzik (left panel) and optimised (right panel) regulator. The horizontal dashed
line indicates the exact value of the ground state energy, whereas the vertical line indicates the value up to which
the numerical integration could be done. The orange dashed line is the extrapolation of our numerical values, given
in blue. In both cases, the ground state energy is obtained to surprisingly high accuracy.}
\label{fig:E0_exppot_regcomp}
\end{figure*}

\subsection{Models}

We will consider three different potentials. As a first example, we will treat
\begin{equation}\label{eq:arctanpot}
 V(x) = \frac{2}{\pi} \arctan \left( x^2 \right) \, .
\end{equation}
This potential carries no additional special properties besides the boundedness.
We include it, because one can solve the flow in a large $N$ approximation
exactly and explicitly for this potential.
As a second potential, we choose a modified version of the well-known P\"oschl-Teller potential,
\begin{equation}\label{eq:poeschl}
 V(x) = \frac{\lambda(1+\lambda)}{2} \left( 1 - \frac{1}{\cosh^2(\lambda x)} \right) \, .
\end{equation}
For this potential, the Schr\"odinger equation can be solved exactly, and all bound states and their
corresponding energies are known \cite{Poeschl1933}. In this work, we will specify to the case $\lambda=1$.
The P\"oschl-Teller potential is also interesting from
another point of view: it is reflectionless for $\lambda \in \mathbbm N$, so waves are transmitted completely through the well.
Lastly, we shall investigate the influence of non-analyticities by studying the potential
\begin{equation}\label{eq:exppot}
 V(x) = e^{-1/x^2} \, .
\end{equation}
All potentials are normalised such that they go to 1 when the argument goes to infinity,
and vanish at their minimum $x=0$.

\subsection{Exact results}

Here we present the (partly numerically) exact solutions for the ground state and
the first excited state (if it exists) for all potentials by solving the 
Schr\"odinger equation (in natural units),
\begin{equation}
 -\frac{1}{2} \Psi''(x) + U(x) \Psi(x) = E \Psi(x) \, .
\end{equation}

For the P\"oschl-Teller potential with $\lambda=1$, there is only
one bound state,
\begin{equation}
 \Psi_0(x) = \frac{1}{\cosh(x)}\, , \qquad \qquad E_0 = 1/2 \, .
\end{equation}

For the other potentials, we apply pseudo-spectral methods along the lines of 
\cite{Borchardt:2015rxa} to obtain the first two bound states. For the potential
\eqref{eq:arctanpot}, the ground state energy, $E_0$, and the energy gap,
$\Delta E = E_1 - E_0$, are
\begin{equation}
 E_0 = 0.448004\, , \qquad \qquad \Delta E = 0.509453 \, .
\end{equation}
On the other hand, for the non-analytic potential \eqref{eq:exppot}, we get
\begin{equation}
 E_0 = 0.356644\, , \qquad \qquad \Delta E = 0.542040 \, .
\end{equation}
All energies and their corresponding wave functions were determined with an accuracy of
at least $10^{-20}$, however there is no need to display more figures in order to
discuss all subsequent results.

\begin{figure*}
\includegraphics[width=\textwidth]{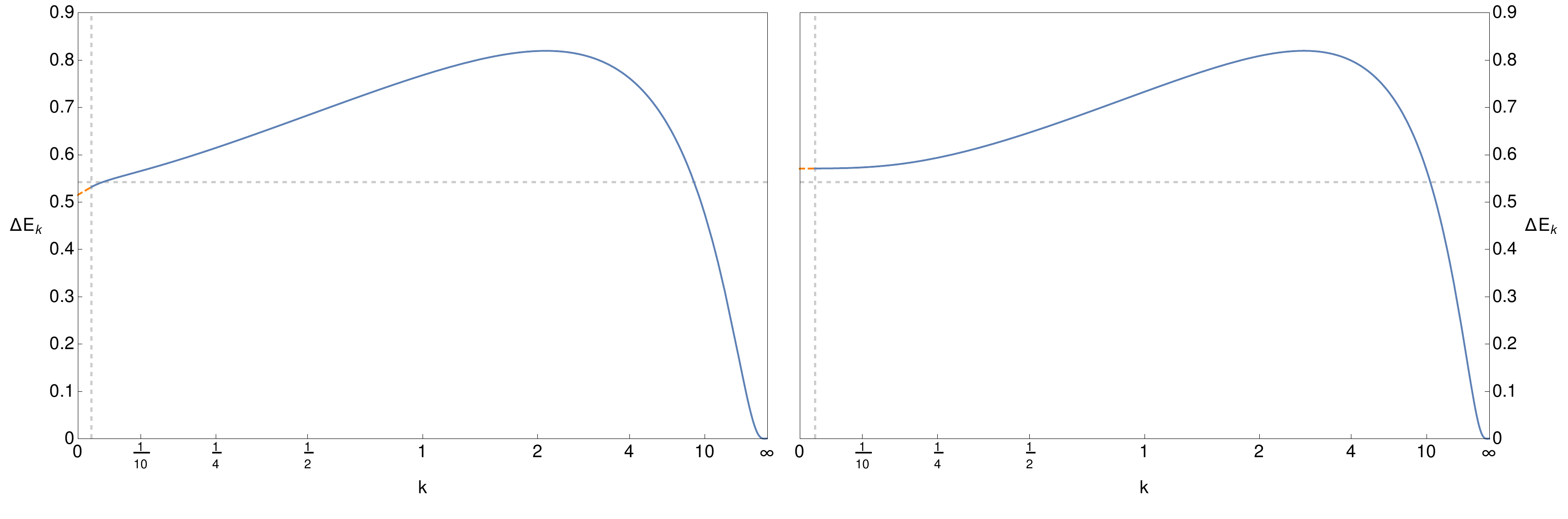}
\caption{Flow of the derivative of the effective potential at vanishing position, which gives the effective energy gap at
scale $k$, $\Delta E_k$, for both Callan-Symanzik (left panel) and optimised (right panel) regulator. The horizontal dashed
line indicates the exact value of the energy gap, whereas the vertical line indicates the value up to which
the numerical integration could be done. The orange dashed line is the extrapolation of our numerical values, given
in blue. The energy gap comes out quite well in both cases.}
\label{fig:DeltaE_exppot_regcomp}
\end{figure*}

\subsection{WKB approximation}

In order to assess the following results, we compare them with the \WKB{} approximation.
The formula for the approximated energy levels reads
\begin{equation} \label{eq:WKB}
 \int_{-x_0}^{x_0} \sqrt{2(E_n-U(x))} = \left(n+\frac{1}{2}\right)\pi,
\end{equation}
where $x_0$ is the classical turning point, $U(x_0)=U(-x_0)=E_n$. 
The index $n$ counts the energy level.
Evaluating \eqref{eq:WKB} for each model, we obtain
for the first potential, \eqref{eq:arctanpot},
\begin{equation}
 E_0 \approx 0.520 \, , \qquad E_1 \approx 0.955 \, , \qquad \Delta E \approx 0.435 \, .
\end{equation}
For the P\"oschl-Teller potential, \eqref{eq:poeschl}, the ground state energy is
\begin{equation}
 E_0 \approx 0.582 \,.
\end{equation}
Finally for the last potential, \eqref{eq:exppot}, we have
\begin{equation}
 E_0 \approx 0.405 \, , \qquad E_1 \approx 0.905 \, , \qquad \Delta E \approx 0.500 \, .
\end{equation}
It is remarkable that $E_1$ deviates less than $1\%$ from the exact value, whereas $E_0$ is off by $13\%-16\%$. 
This is to be expected, since the \WKB{} approximation works well in the semiclassical limit $\lambda \ll 2x_0$,
where $\lambda/2$ is the distance between two knots of the wave function.
This translates into the condition $n \gg 1$.

\subsection{1-loop approximation}

As a further step to put subsequent results in perspective, we perform a 1-loop calculation.
The 1-loop effective potential reads
\begin{equation}
 U_\text{1-loop}(x) = U_\text{cl}(x) + \frac{1}{2} \sqrt{U_\text{cl}''(x)} \, ,
\end{equation}
which can for example be obtained directly from the flow equation \eqref{eq:Wetterich} by setting the potential on the 
right-hand side equal to the classical potential $U_\text{cl}$. The ground state energy is
given by the value of the effective potential at its minimum (here in all cases $x=0$), whereas
the energy gap is the square root of the curvature of it, also evaluated at the minimum.
One thus obtains for the first potential, \eqref{eq:arctanpot},
\begin{equation}
 E_0 = \frac{1}{\sqrt{\pi}} \approx 0.564 \, , \qquad \Delta E = \frac{2}{\sqrt{\pi}} \approx 1.128 \, .
\end{equation}
The ground state energy comes out more or less well for such a simple calculation, but the 1-loop
result predicts that there are no further bound states, as the energy gap is too large.

For the P\"oschl-Teller potential, the 1-loop result is
\begin{equation}
 E_0 = \frac{1}{\sqrt{2}} \approx 0.707 \, ,\, \Delta E = \sqrt{2\left( 1-\sqrt{2} \right)} \approx 0.910 \mathbf{i}\, .
\end{equation}
The convexity of the effective potential is not caught by a 1-loop calculation, and accordingly, the 
energy gap is imaginary. This phenomenon is well-known to be an artifact of the loop expansion, and extensively
discussed in {\it{e.g.}} \cite{Fujimoto:1982tc,Coleman:1973jx}. The ground state energy is off by about 40\%.

Finally, for the non-analytic potential \eqref{eq:exppot}, no meaningful 1-loop analysis can be done.
In fact, any order in perturbation theory fails to produce anything non-zero for the energy levels
because of the non-analyticity.

\begin{table}
\begin{tabular}{|c|c|c|c|}
\hline
 \multicolumn{4}{|c|}{$V(x) = 2/\pi \arctan \left( x^2 \right)$}\\ \hline
  \qquad\qquad\qquad& \qquad\, exact\qquad\, & \qquad\,{\small{CS}}\qquad\, & \qquad\,opt\qquad\, \\ \hline
  $E_0$ & 0.448004 & 0.445 & 0.447 \\
  $\Delta E$ & 0.509453 & 0.477 & 0.558 \\ \hline
 \multicolumn{4}{|c|}{$V(x) =  1 - 1/\cosh^2(x)$}\\ \hline
  & exact & {\small{CS}} & opt \\ \hline
  $E_0$ & 1/2 & 0.496 & 0.499 \\ 
  $\Delta E$ & - & 0.464 & 0.585 \\ \hline
 \multicolumn{4}{|c|}{$V(x) = \exp(-1/x^2)$}\\ \hline
  & exact & {\small{CS}} & opt \\ \hline
  $E_0$ & 0.356644 & 0.355 & 0.356 \\ 
  $\Delta E$ & 0.542040 & 0.515 & 0.570 \\ \hline
\end{tabular}
\caption{Overview of exact results from solving the Schr\"odinger equation
and results obtained from the flow of the potential for all three potentials.
{\small{CS}} and opt indicate that the Callan-Symanzik and the optimised regulator were
employed, respectively.}
\label{tab:results}
\end{table}

\subsection{Flow of the effective potential}

This section is devoted to the numerical study of the actual flow equation for the effective potential.
All investigations are done within \LPA{} where $Z\equiv 1$.
Note, that in quantum mechanics no renormalisation is needed.
Therefore, the initial condition can be put at $k=\Lambda\rightarrow \infty$.
To cover the whole interval $k\in[0,\infty)$ the time direction is compactified analogously to \eqref{eq:compactification}.
As in the previous section, for reasons of numerical stability, we actually use the flow equation for the derivative of the effective
potential $U'(\rho)=\partial_{\rho}U(\rho)$, and obtain the ground state energy by an additional integration.
The flow equation reads
\begin{equation} \label{eq:flowQM}
 \partial_k U'(\rho) = -A k^B \frac{3 U''(\rho)+2\rho U'''(\rho)}{(k^2 + U'(\rho)+2\rho U''(\rho))^C},
\end{equation}
where $A=1/\pi, \, B=C=2$ for the linear optimised regulator ($R_k(p^2) = (k^2-p^2)\theta(k^2-p^2)$) and
$A=1/4, \, B=1$ and $C=3/2$ for the Callan-Symanzik cutoff ($R_k(p^2) = k^2$).

We will first point out some expectations on the outcome of the flow, followed by the discussion of
the actual results of the flow. An overview of all results can be found in \autoref{tab:results}.\\

{\it{Expectations - }}The effective potential needs to be convex at $k=0$ (except in particular cases, see
the discussion in the next section). It is immediately clear that any bounded function that is not constant
cannot be convex. It follows that if we could integrate the flow equations down to $k=0$, we would end up
with a constant potential, and the constant is exactly the ground state energy.
One can prove this by considering an alternative definition of the effective potential \cite{Curtright:1983cf},
\begin{equation}
 U(\bar x) = \inf\limits_{\Psi: \langle x \rangle = \bar x} \langle H \rangle \, ,
\end{equation}
that is, the effective potential at a point $\bar x$ is given by the infimum of the Hamiltonian over all states
with position expectation value $\bar x$. Exhaustive
discussions of the effective potential in quantum field theory can be found in {\it{e.g.}} \cite{9780511565045,Curtright:1983cf,
Symanzik:1969ek,Stevenson:1984rt}.
Our naive expectation on the flow is therefore that we can hope to find the ground state energy, but probably not 
the energy of the first excited state. Surprisingly, it turns out that one can extract some estimate of the 
excited state energy from the flow.\\

{\it{Numerical results - }}As exemplary case, we display the numerical results from solving the flow equation
for the non-analytical potential \eqref{eq:exppot}. The other two potentials pose no further challenge and show
the same qualitative behaviour.

\begin{figure}
\includegraphics[width=\columnwidth]{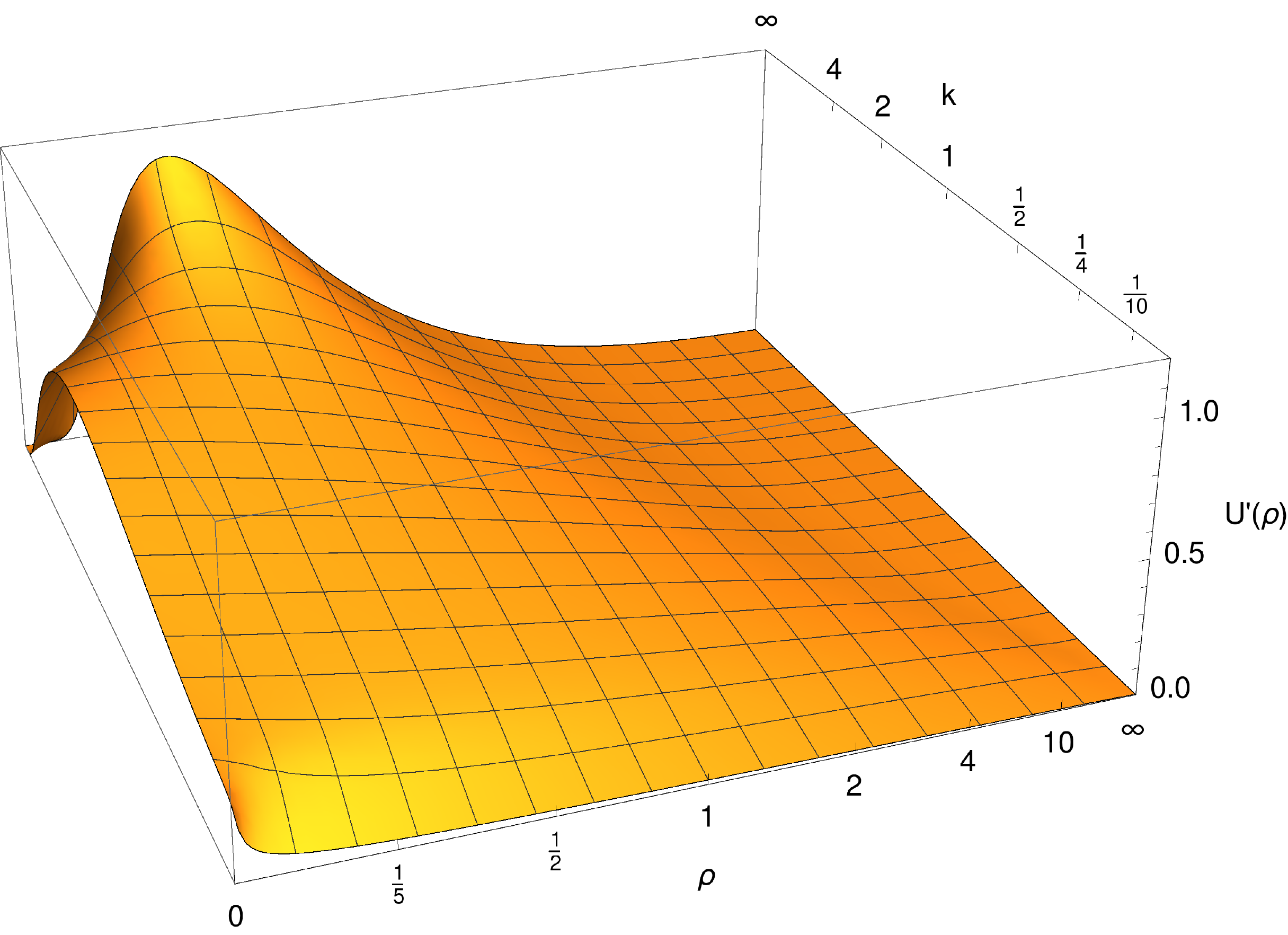}
\caption{Flow of the derivative of the effective potential for the Callan-Symanzik regulator. One can see that
the non-analyticity of the classical potential quickly smooths out. Convexity problems for small scales
arise for large values of the position, in contrast to conventional unbounded potentials.}
\label{fig:exppot_CS_flow}
\end{figure}

In \autoref{fig:E0_exppot_regcomp}, the effective potential at $x=0$ as a function of the scale $k$ is depicted, for both
the Callan-Symanzik and the optimised regulator. It corresponds
to the effective ground state energy at scale $k$. The horizontal dashed line indicates the exact value obtained from
the Schr\"odinger equation. For technical reasons, we cannot integrate down to $k=0$, but only to a finite value, indicated
by the vertical dashed line. From thereon, we extrapolate linearly to get an estimate of the true ground state energy.
For both regulators, we get very precise estimates for the ground state energy. Generically, the optimised regulator gives
slightly superior results for $E_0$.

Next, we shall discuss the results on the energy gap. As argued above, in principle we should not expect to get any
meaningful estimate from the effective potential. There is however a loophole in the above argument: it is based
on the effective potential at scale $k=0$, when all fluctuations are integrated out. When we consider the flow of the
effective potential, we can extract further information, as the scale $k$ is roughly the (inverse) scale of a finite box
that the system lives in, giving an effective cutoff to physics. In this sense, we can indeed extract information on the energy gap,
roughly when the scale is large enough to resolve the wave function of the first excited state, but small enough not to be too much influenced
by the next-higher states. Bearing this in mind,
we shall discuss the first derivative of the flowing potential, again at vanishing position, which gives the effective
energy gap at scale $k$ \cite{Kapoyannis:2000sp},
\begin{equation}
 \Delta E = \sqrt{U'(\rho)}\,|_{\rho=0} \, ,
\end{equation}
It is shown in \autoref{fig:DeltaE_exppot_regcomp}, again for both regulators.
Remarkably, in both cases again, we get a quite good estimate of the true energy gap, however the finer details are more complicated.
For the Callan-Symanzik regulator, we can already see the influence of convexity, as the derivative of the effective potential bends
towards zero. This is not the case for the optimised regulator yet. Correspondingly, the optimised regulator overestimates
the energy gap, whereas the estimate from the Callan-Symanzik regulator is below the true value. This behaviour is also observed for
the other potentials, and influences the prediction of the number of bound states. In this respect, the optimised regulator
erroneously predicts only one bound state for the potential \eqref{eq:arctanpot}. On the other hand, the Callan-Symanzik regulator
predicts a second bound state for the P\"oschl-Teller potential \eqref{eq:poeschl}. Either way, any prediction for the energy
gap from the flow should be taken with a grain of salt, as convexity has to set in at some point, and also the extrapolation
introduces further errors. Presumably one should read off the energy gap at some finite value of the scale, at which the
first excited state is completely resolved, however we found no a priori argument on how to set this scale.

In \autoref{fig:exppot_CS_flow}, we depict the actual flow of the derivative of the effective potential,
obtained with the Callan-Symanzik regulator. One can see that
the non-analyticity of the classical potential is smoothed out quickly. For small scales $k$, one can also see the tendency
of the derivative of the effective potential to flow to zero, as it must due to convexity. In contrast to
unbounded potentials, where convexity is numerically challenging near the origin, the numerical problems here arise for large values of the
position, which makes it increasingly difficult to resolve the flow.

\begin{figure}
\includegraphics[width=\columnwidth]{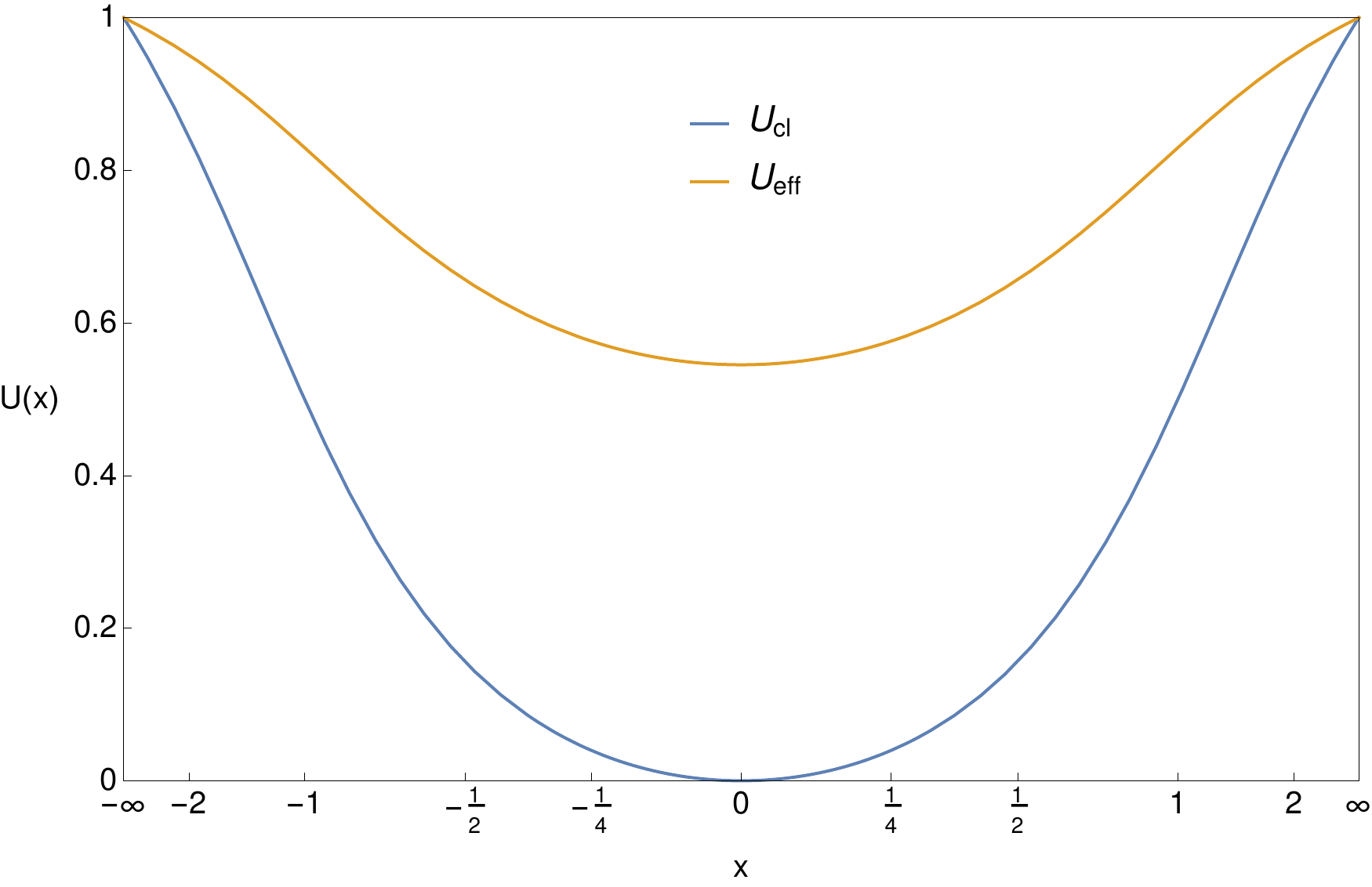}
\caption{Comparison of the classical potential \eqref{eq:arctanpot} and the corresponding effective potential \eqref{eq:arctanlargeN}
in the large $N$ limit. In contrast to finite $N$, we do not observe convexity of the effective potential,
but only that the $\rho$-derivative is non-negative.}
\label{fig:QM_largeN}
\end{figure}

\subsection{Large \texorpdfstring{$N$}{N} approximation}

As a final point, we shall study the potential \eqref{eq:arctanpot} in the limit of infinitely many dimensions,
similar to a large $N$ approximation in the O$(N)$-model. This means specifically that the index $a$ in \eqref{eq:effaction} counts the space coordinates,
and we allow it to run from 1 to $N$, sending $N\to\infty$.
In this case, the flow equation can be
solved implicitly by the method of characteristics \cite{Tetradis:1995br}. In the case of the potential \eqref{eq:arctanpot}, the implicit
relation $x=x(U)$ can be inverted, delivering the full effective potential. It is given by
\begin{equation}\label{eq:arctanlargeN}
\begin{aligned}
U(x) &= \frac{-\pi x^2 + \sqrt{16\pi(1+x^4)-\pi^2}}{8\pi(1+x^4)} \\
 &+ \frac{2}{\pi} \left( \arctan(x^2) + \arctan\left( \sqrt{\frac{\pi}{16(1+x^4)-\pi}} \right) \right).
\end{aligned}
\end{equation}
Notably, the large $N$ effective potential is {\it{not}} convex. This seeming paradox has the following reason.
Convexity is tied to the condition that the propagator avoids a singularity for negative $U''(x)$ which appears
in the equivalent of the radial mode propagator. In the large $N$ approximation, however,
only the equivalent of the Goldstone mode propagator survives, and for it to be
finite, it is enough that $U'(\rho) \equiv U'(x)/x$ is non-negative. This is indeed the case for the solution given above.
A plot of both the classical and the effective potential is given in \autoref{fig:QM_largeN}.

\section{Summary}\label{sec:sum}

We extended the ideas from previous work on functional fixed point equations to also
solve functional flow equations to high accuracy.
We first discussed flows of the O$(N)$-model
in three dimensions, for $N=1,\,4$ and in the large $N$ limit.
In all cases, we could achieve a highly stable
and precise flow.
We showed that our method can accomplish the time integration to
machine precision, and always stays very close to the analytical solution exactly known in the large $N$ limit. The error
in this case is dominated by the condition of the differential equation.
Even for numerically challenging tasks, as resolving the convexity of the effective potential in the \IR{},
the flow was traceable for 6 orders of magnitude for $N=4$, and about 2 orders of
magnitude for $N=1$.
Then, we calculated the flow along a separatrix from the first multicritical fixed point to the Wilson-Fisher
fixed point in $d=2.4$, for which almost 13 orders of magnitude were integrated out at high precision.
As a second model, we treated a set of bounded potentials in $d=1$, which are reminiscent to potentials in a quantum field theory context such as Higgs inflation.
Technically, they are interesting because they need global resolution for a numerically stable flow.
For the three potentials that we discussed, we extracted the ground state and first excited state
energies in an \LPA{} truncation to satisfying accuracy, even though one might have expected from
analytic arguments that the determination of the first excited state energy was not possible
from the effective potential alone. Finally, also non-analyticities pose no problem to our
method, in contrast to expansions in powers of the field.

\section*{Acknowledgements}

We would like to thank 
A. Bonnano, H. Gies and A. Wipf for useful discussion and
H. Gies and A. Wipf for valuable comments on the manuscript.
This work was supported by the DFG-Research Training Group
``Quantum and Gravitational Fields'' GRK 1523/2.
J.B. acknowledges further support by DFG under grant no. Gi 328/6-2 (FOR 723), while
B.K. is grateful for funding by DFG grant no. Wi 777/11-1.

\bibliography{specbib}

\end{document}